\documentclass[aps,pre,twocolumn,amsmath,floatfix,showpacs,superscriptaddress]{revtex4}

\usepackage{graphics}
\usepackage{amsmath}
\usepackage{epsfig}
\usepackage{bm}

\begin{document}

\title{Asymptotic boundary layer method for unstable trajectories: Semiclassics for individual scar wavefunctions}

\author{A. Vagov}
\affiliation{Institut f\"ur Theoretische Physik III, Bayreuth Universit\"at,
Bayreuth 95440, Germany}
\affiliation{Department of Physics, Lancaster University,
Lancaster LA1 4YB, United Kingdom}
\author{H. Schomerus}
\affiliation{Department of Physics, Lancaster University,
 Lancaster LA1 4YB, United Kingdom}
\author{V. V. Zalipaev}
\affiliation{Steklov Mathematical Institute, Fontanka 27, St.-Petersburg 191023, Russia}
\affiliation{ Department of Mathematical Sciences, University of
Liverpool, Liverpool L69 7ZL, United Kingdom}

\date\today

\begin{abstract}
We extend the asymptotic boundary layer (ABL) method, originally
developed for stable resonator modes, to the description of
individual wavefunctions localized around unstable periodic
orbits.  The formalism applies to the description of scar states
in fully or partially chaotic
 quantum systems, and also allows for the presence of smooth and
sharp potentials, as well as magnetic fields. We argue that the
separatrix wave function provides the largest contribution to the
scars on a single wave function.  This agrees with earlier results
on the wave-function asymptotics and on the quantization condition
of the scar states. Predictions of the ABL formalism are compared
with the exact numerical solution for a strip resonator with a
parabolic confinement potential and a magnetic field.
\end{abstract}

 \maketitle

\section{Introduction}

Semiclassical methods provide valuable insight into the eigenmodes
of electromagnetic microresonators and mesoscopic electronic
devices because they establish a direct relation to classical
trajectories.  For stable orbits the corresponding wavefunctions
are localized in the vicinity of the trajectory and can be
constructed by the elegant asymptotic boundary layer (ABL) method,
which is based on rectified Gaussian beams and their harmonic
transverse excitations \cite{BaKi,BaBu,Bel,Zalipaev}.  Describing
wavefunctions associated with unstable orbits is a considerably
more difficult problem.  For classically chaotic quantum systems
\cite{Stock}, the rapid stretching and folding of the phase space
translates into a randomization of most eigenstates, which can be
addressed through statistical descriptions \cite{Berry2,Voros}.
However, some eigenstates deviate significantly from a random wave
pattern in that they are enhanced in the vicinity of unstable
orbits \cite{Heller,Kaplan,Stock}.  These so-called scars are most
pronounced for short periodic orbits, and often dominate the
individual properties of quantum systems.  Their influence has
been detected in numerous numerical investigations and experiments
on a large variety of systems, such as microwave cavities
\cite{microwave}, resonant tunnelling diodes \cite{tunneling},
and, in particular, in the lasing modes of dielectric
microresobators \cite{microlasers}.

Pioneering theoretical investigations of individual scar
wavefunctions retained the assumption of a Gaussian transverse
profile and its harmonic excitations
\cite{Heller,Kaplan,Antonsen,Dambrowsky}.  It was soon realized
that this requires to superpose a large number of transverse
excitations \cite{Vergini,keating2009}.  On average, the
large-distance asymptotic decay of the transverse profile is
expected to follow a power law
\cite{Bogomolny,Berry,Agam,Kaplan2}.  Studies of hyperbolic fixed
points (with effective period 0) also point into the direction
that the transverse profiles are essentially non-Gaussian
\cite{Nonnenmacher,Mendes,Lee}.  On the other hand, numerical
investigations \cite{Fromhold} have led to the conjecture that
scar quantization only involves a single (longitudinal) quantum
number, which means that the transverse mode profile is fixed.

In this paper we clarify the nature of scarred states from the
perspective the ABL method, which offers a systematic
semiclassical expansion for the Schr\"odinger equation in the
asymptotic boundary layer around a classical trajectory
\cite{BaKi}. Compared to other semiclassical methods, a main
advantage of the ABL method is its applicability to a broad class
systems, including systems in external potentials  and magnetic
fields. ABL expansion up to quadratic terms of the effective
potential yields the Gaussian wave packet solution, originally
proposed by Heller \cite{Heller} as a convenient ansatz to study
dynamics of complex molecules. Subsequent mathematical studies
\cite{Mas1,BaKi,BaBu} advanced the method into a powerful general
tool to obtain asymptotic solutions to various equations of
theoretical and mathematical physics, ranging from optics to
gravity.

Most of these applications, however, focus on stationary
eigenstates associated with stable periodic orbits. The
application of the ABL formalism to \emph{unstable} orbits faces
the problem that Gaussian beams fail to satisfy the periodicity
condition along such orbits. In principle, this problem can be
circumvented by numerical superposition of a large (ideally,
infinite) number of Gaussian solutions for an artificially
stabilized orbit \cite{Vergini}. However, as we will show, it is
possible to take a more direct analytic approach, which focusses
on the mathematical difference of the ABL equations for stable and
unstable trajectories. This allows to obtain closed analytical
expressions for individual wavefunctions localized around unstable
trajectories. A detailed analysis of these solutions reveals the
special role of the so-called separatrix solution in the formation
of scar states. This  yields a transverse mode profile that
corresponds to an appropriately rectified specific solution of the
hyperbolic fixed-point problem \cite{Nonnenmacher,Mendes,Lee} and
also satisfies the correct large-distance asymptotics
\cite{Bogomolny,Berry}.  The periodicity condition for the
separatrix solution leads to a Bohr-Sommerfeld quantization
formulae with a single longitudinal quantum number, in agreement
with the earlier numerical conjecture of Ref.\ \cite{Fromhold}.
These general considerations are presented in Sections
\ref{sec2}-\ref{sec4}.

In Section  \ref{sec5} we illustrate the flexibility of the ABL
approach by applying it to the example of a two-dimensional strip
resonator in a magnetic field and a parabolic confinement
potential. Comparison of the ABL predictions with the results of
numerical computations demonstrates excellent accuracy of the
method.

Section \ref{sec6} contains discussion and conclusions. Details of
the derivations are presented in the Appendix.

\section{\label{sec2}ABL formalism}

In the following two sections we provide the main results of the
ABL formalism and cast them into an outfit which naturally
contrasts states localized around stable and unstable
trajectories. Details of the derivation of the main formulas are
given in Appendix. To make the presentation compact we concentrate
on two-dimensional quantum systems in a constant perpendicular
magnetic field $B$, with vector potential ${\bf A} = B (-y,x)/2$.

We employ a curvilinear system of coordinates associated with the
main classical trajectory ${\bf r}_{0}(s)$, which is parameterized
by its arclength $s$. A pair of coordinates $(s,n)$ defines the
position vector in the vicinity of the trajectory as ${\bf r} =
{\bf r}_{0}(s) + {\bf e}_n(s) n$, where ${\bf e}_n$ is the normal
unit vector and $n$ measures the distance from the trajectory. The
ABL formalism seeks a solution in an asymptotically small boundary
layer of width $\propto \sqrt{\hbar}$ around the main trajectory,
which introduces a natural scaling of the distance as $n =
\sqrt\hbar\, \nu$. The solution of the Schr\"odinger equation is
sought in the form
\begin{equation}
    \Psi(s,\nu) = e^{\frac{i}{\hbar }
    (S_{0}(s) + S_{1}(s)  \sqrt\hbar\, \nu)} \sum_{j=0}^{\infty}
\hbar^{j/2}
    \psi^{(j)}(s,\nu),
    \label{eq:WKB}
\end{equation}
which represents an expansion in {\it half-integer} powers of
Planck's constant $\hbar$ (in contrast to the text-book WKB
expansion in integer powers of $\hbar$). This expansion of the Schr\"odinger equation (see Appendix) yields the exponent in Eq.\
(\ref{eq:WKB}) in the form of the classical action functions
\begin{subequations}
\label{eq:S}
\begin{eqnarray}
      &&S_{0} = \int_{0}^{s} \left\{ a(s) + e(A_{x} e_{n}^{y}
      - A_{y} e_{n}^{x})  \right\} ds,   \\
      && S_{1} =
e ( A_{x} e_{n}^{x} + A_{y} e_{n}^{y}) , \\
      \label{eq:a}
&&
      a(s) = \sqrt{2m[E - u({\bf r}_{0}(s))]},
\end{eqnarray}
\end{subequations}
where $u({\bf r})$ is the scalar potential. For $\psi^{(j)}$ one
obtains a hierarchical set of equations. If the expansion of the
effective potential in the system can be restricted to the terms
of   second order in $n$ (the standard assumption in the linear
theories \cite{Kaplan,Dambrowsky}),
 one can restrict the series in Eq.\ (\ref{eq:WKB})
to the leading term $\psi\equiv \psi^{(0)}$ and obtains the
so-called boundary layer Schr\"odinger equation
\begin{eqnarray}
\label{eq:SL}
    && i {\dot \psi}(s,\nu) = \left(- \frac{1}{2 a}
    \partial_{\nu}^{2} + \frac{a d}{2}
    \nu^{2} -  \frac{i}{2} a \right)
    \psi(s,\nu), \\
    &&
    d(s) = \frac{2 m u_{2}}{a^{2}} + \frac{m^2
u_{1}^{2}}{a^{4}} -
    \frac{2 m u_{1}}{\rho a^{2}} - \frac{eB}{\rho a}. \nonumber
\end{eqnarray}
Here the dot denotes the partial derivative $\partial_{s}$, while
$\rho$ is the geometric radius of curvature of the trajectory. The
functions $u_{1,2}$ are obtained from the expansion
\begin{align}
u({\bf r}) \approx u_{0}({\bf r}_{0}(s)) + u_{1}(s) \sqrt{\hbar}\,
\nu + u_{2}(s) \hbar\, \nu^{2} .
\end{align}

Equation (\ref{eq:SL}) is analogous to the Schr{\"o}dinger
equation for a one-dimensional {\em nonstationary} oscillator,
where $s$, $a(s)$, and $d(s)$ take the role of time,  mass, and
harmonic frequency, respectively. As shown in the Appendix, a
general form of its partial solutions can be constructed by
establishing their relation with the classical trajectories in the
vicinity of the main one. The final result yields a set of
solutions of the form
\begin{equation}
    \label{eq:psi}
    \psi_{\xi}(s,\nu) \propto
    \frac{ {\overline z}^\xi (\Gamma-{\overline \Gamma})^{\xi/2}}
     {\sqrt{a z}}
     D_{\xi} \left( \sqrt{\frac{\Gamma - {\overline \Gamma}}{2i}} ~\nu
     \right)
          e^{ \frac{i}{4} (\Gamma+ \overline \Gamma) \nu ^{2} },
\end{equation}
where $D_{\xi}$ are parabolic cylinder functions with arbitrary
index $\xi$, while $\Gamma = p/z$ and $\overline \Gamma = \overline p /\overline z$ are defined by two independent solutions of the Hamilton {\em equations in variation}
\begin{equation}
    {\dot  z} = \frac{p}{a}, \quad{\dot p}  = - a d z,
    \label{eq:HE}
\end{equation}
that satisfy $p {\overline z} - {\overline p} z = w$  ($w$ is the
Wronskian, which here is a constant).  A pair $(z, p)$ describes
the classical trajectories in the vicinity of the main trajectory.
Once Eqs. (\ref{eq:HE}) are solved, Eqs. (\ref{eq:WKB}),
(\ref{eq:S}) and (\ref{eq:psi}) define semiclassical solutions of
the Schr{\"o}dinger equation. A second class of partial solutions
is found by interchanging $z \leftrightarrow {\overline z}$ in
Eq.\ (\ref{eq:psi}).

The similarity of Eq.\ (\ref{eq:psi}) with a standard oscillator
can be seen by assuming constant $a$ (``mass'' $m$) and $d$
(squared ``frequency'' $\omega^2$). If one also chooses solutions
to Eq.\ (\ref{eq:HE}) as $z \propto \exp (i \omega s)$,  Eq.\
(\ref{eq:psi}) yields familiar oscillator wave functions. In the
general case, the detailed dependence of $a$ and $d$ on $s$
results in nontrivial modulations of the wave function profile
along the trajectory.

\section{Periodic orbits}

Since there are two partial solutions of Eq.\ (\ref{eq:HE}), the
general solution Eq.\ (\ref{eq:psi}) for the wave function
possesses one degree of freedom. For closed periodic orbits,
however, this freedom is restricted by an additional periodicity
condition. As we shall see, this condition naturally distinguishes
between stable and unstable orbits.

For stable orbits we follow a standard procedure, the first step
of which is to find the Floquet solutions of Eq.\ (\ref{eq:HE}),
which fulfill the conditions $z(T)= \exp (i \phi) z(0)$ and $p(T)=
\exp (i \phi) p(0)$. Here $T$ is the period of the orbit and
$\phi$ is a real Floquet phase. If $(z_{1,2},p_{1,2})$ are two
independent solutions with initial conditions $(1,0)$ and $(0,1)$,
respectively, the general solution is a linear combination
\begin{align}
 \left( \begin{array}{c} z \\ p \end{array} \right) = \alpha_1 \left(
 \begin{array}{c} z_1 \\ p_1 \end{array} \right)+ \alpha_2 \left(
 \begin{array}{c} z_2 \\ p_2\end{array} \right).
\end{align}
Floquet solutions are found by determining $\alpha_1$, $\alpha_2$,
and $\Lambda = \exp (i \phi)$ from the eigenvalue problem
\begin{align}
 M \left( \begin{array}{c} \alpha_1 \\ \alpha_2 \end{array} \right) =
 \Lambda \left( \begin{array}{c} \alpha_1 \\ \alpha_2 \end{array} \right),~
 M= \left( \begin{array}{cc} z_1(T) & z_2 (T) \\ p_1(T) & p_2 (T)
 \end{array} \right).
\end{align}
Here $M$ is the monodromy matrix (with a unit determinant), which
depends on the stability of the orbit.

For stable orbits $|{\rm Tr}\ M|<2$, and $M$ has two complex
conjugate eigenvalues with a unit absolute value. The
corresponding Floquet solutions are also complex conjugate,
$(z,p)$ and $({\overline z},{\overline p}) = (z^*,p^*)$. We use
the convention that $(z,p)$ is the solution with ${\rm
Im}\,\Gamma>0$.  This  choice is well-defined since ${\rm
Im}\,\Gamma \propto |z|^{-2} $  does not change sign along the
trajectory. With this choice, Eq.\ (\ref{eq:psi}) is restricted to
normalizable harmonic oscillator functions, which are also
referred to as the Gaussian beam (or ray) solutions
\cite{BaBu,BaKi}. The periodicity condition for the solution
defined by Eqs.\ (\ref{eq:WKB}), (\ref{eq:S}) and (\ref{eq:psi})
is obtained from the phase increment for a single round trip along
the orbit, which depends on the classical action $S_0$ and the
phases of the various powers of $z,\overline z$ in the
expressions.  This yields the Bohr-Sommerfeld quantization formula
for stable orbits,
\begin{equation}
\label{eq:quant}
   \int_{0}^{T} a(s) ds + \Phi = \hbar[2 \pi n +\left( m +1/2
   \right) \phi(E)],
\end{equation}
where $\Phi$ is the magnetic flux through the orbit, while $n$ and
$m$ are longitudinal and transverse quantum numbers.

For unstable orbits, $|{\rm Tr}\ M| > 2$. In this case the
eigenvectors and eigenvalues of $M$ are real, such that
$\Lambda_\pm = \exp (\pm \lambda)$ where $\lambda$ is the
dimensionless Lyapunov exponent (corresponding to a time-domain
Lyapunov exponent $\lambda_T=\lambda/T$). For this case the
harmonic oscillator functions with integer index $m$ in Eq.\
(\ref{eq:psi}) violate the periodicity condition. However, the
periodicity can still be achieved with the choice of the $\xi =
-1/2 + i \eta$ (where $\eta $ is real).  The resulting functions
are similar to those for an inverted (negative) harmonic
potential.

The periodicity requirement of the wave function then results in
the Bohr-Sommerfeld like quantization condition for unstable
trajectories
\begin{equation}
\label{eq:quant-unst}
    \int_{0}^{T} a(s) ds + \Phi =\hbar[2 \pi n \pm  \eta \lambda(E) + \pi
    \alpha/2].
\end{equation}
The Maslov index  $\alpha$ counts   singular points along the
trajectory, where $z$ or $\overline z$ vanishes. Since $\Gamma
\propto z^{{-1}}$, these points produce a square-integrable
singularity $\propto z^{-1/4}$ in the wave function [see also Eq.
(\ref{eq:psi-unst}), below], which can be further regularized by
uniform approximations (see, {\em e.\,g.}, Refs.
\cite{Berry3,bifurcations}).

\section{\label{sec4}Contribution to scar states}
The ABL solutions defined by Eqs.\ (\ref{eq:psi}) and
(\ref{eq:HE}) are only valid  in a boundary layer of width $|n|
=\sqrt{\hbar} |\nu| = O(\sqrt{\hbar})$ around the orbit. In order
to construct eigenstates associated with the unstable trajectory,
the semiclassical wave function has to be matched to a
quasi-random background  beyond this layer \cite{nonnenmacher2},
which in principle can mix  solutions with different values of
$\eta$ and $n$. The number of contributing solutions is defined by
the interplay between the coupling of the solutions with the
background, and the energetics of the different states as
determined by the quantization condition in Eq.\
(\ref{eq:quant-unst}). In essence, one therefore deals with a
scattering problem. As confirmed in recent works on quantum
resonance wave functions
\cite{schomerustworzydlo,keating2007,keating2009}, an important
scale in this context is the Ehrenfest time \cite{ehrenfest},
defined by the time of a wavepacket of initial size of de Broglie
wave length to spread across the entire accessible phase space of
the system,
\begin{align}
 t_{\rm Ehr} = \lambda_T^{-1} \ln N_{\rm ph}.
\end{align}
Here $\lambda_T=\lambda/T$ is the time-domain Lyapunov exponent,
and  $N_{\rm ph} = L \sqrt{2mE} /\hbar \gg 1$ is a dimensionless
measure (in units of Planck's constant) of the total accessible
volume of the phase space, where $L$ is the characteristic size of
the system.

The Ehrenfest time defines an energy window
\begin{align}
\label{eq:En0}
E = E_{n_0} \pm  \frac{2\pi \hbar} { t_{\rm Ehr} },
\end{align}
over which partial ABL solutions are strongly mixed among each
other via the background states. Here $E_{n_0} = E_{n_0,\eta=0}$
the characteristic energy following from the Bohr-Sommerfeld
quantization condition (\ref{eq:quant-unst}), with $\eta=0$.
Earlier  numerical observations provide evidence that scars are
quantized with energies close to $E_{n_0}$ \cite{Fromhold}. This
is the energy of the symmetric separatrix solution,
\begin{equation}
    \label{eq:psi-unst}
    \psi(s,\nu)\propto \sqrt{ \frac{ \nu}{z {\overline z}}} \
    J_{-1/4}\left(\frac{w \nu^2}{4 z {\overline z}} \right)
    e^{\frac{i}{4} (\Gamma+\bar \Gamma) \nu^{2}} .
\end{equation}
Separatrix solutions without trajectory-specific $z$ and
$\overline z$ were earlier discussed in the context of scarring
phenomenon as the solution of the hyperbolic fixed-point problem
\cite{Nonnenmacher,Mendes,Lee}. The asymptotic form of Eq.\
(\ref{eq:psi-unst}) for large $\nu$
\begin{equation}
    \label{eq:psi-as}
\psi(s, \nu) \propto \frac{1}{\sqrt{\nu}} \cos\left(\frac{w
\nu^2}{4 z{\overline z}} \right),
\end{equation}
recovers the well known  result of Bogomolny \cite{Bogomolny}
(again, generalized to contain the  trajectory-specific $z$),
which is often interpreted as the profile of a scarred wave
function associated with few closely located unstable orbits
\cite{Stock}. We also note that the description of scars in
earlier works \cite{Vergini} employed the ABL formalism for stable
orbits together with the variational procedure to minimize the
squared transverse energy. This effectively determines the wave
function closest to the separatrix solution (\ref{eq:psi-unst}),
but represents this as a sum of harmonic oscillator functions.

Expanding the quantization condition (\ref{eq:quant-unst}) around
$E=E_{n_0}$ one obtains
\begin{align}
\label{eq:quant-app}
E \approx E_{n_0} + \frac{2\pi \hbar}{T}\delta n + \hbar \lambda_T \eta,
\end{align}
where $\delta n = n - n_0$. The approximate 'level spacing' of
states with different $\eta$ can be estimated from the asymptotics
of Eq.\ (\ref{eq:psi}) at $\nu \rightarrow \infty$,
\begin{align}
\label{eq:psi-as0}
 \psi_\eta (s,\nu ) \propto \frac{1}{\sqrt{\nu}} \exp \left\{ i
 \frac{\Gamma }{2} \nu^2 - i \eta \ln \left( \nu \sqrt{\frac{\Gamma - \overline \Gamma}{2}}
 \right) \right\}.
\end{align}
Assuming that the mixing occurs when the state spreads to a
transverse distance $L_{\rm tr}=O(L)$, the typical spacing of
solutions with different $\eta$ is
\begin{align}
\label{eq:eta-quant} \Delta \eta \propto \ln \left(  L_{\rm tr}
\sqrt{\frac{\langle \Gamma \rangle_s - \langle  \overline \Gamma
\rangle_s }{2\hbar} }\right),
\end{align}
where $\langle \rangle_s$ denotes averaging over the trajectory.

Under certain conditions,  only a single solution (\ref{eq:psi})
is found in the Ehrenfest window (\ref{eq:En0}). From the
longitudinal quantization one obtains the standard condition
\begin{equation}
T \lesssim t_{\rm Ehr} . \label{eq:tvstehr}
\end{equation}
The  transverse quantization gap (\ref{eq:eta-quant}) implies
\begin{align}
\label{eq:cond-tr-f} T \lesssim  O ({\lambda_T}^{-1} N_{\rm
ph}^\zeta),
\end{align}
with $\zeta>0$ depending on the precise choice of $L_{\rm tr}$.
This condition is satisfied for all $T \lesssim t_{\rm Ehr}$, and
therefore is weaker than the longitudinal quantization condition.
Consequently, as long as the period of the orbit $T$ is less than
the Ehrenfest time, only a single ABL solution will contribute to
the semiclassical eigenstate.

The arguments from the above procedure of matching of the ABL
solution to the random background therefore recovers the main
phenomenological result on the strong scarring along unstable
trajectories, Eq.\ (\ref{eq:tvstehr}), which was earlier obtained
by analysis of time-dependent Gaussian packets \cite{Heller3}.
Because of the conceptual relation to a scattering problem, this
argumentation can be enforced by taking the transverse probability
flux across the scar into account. This can be studied using the
probability current in  curvilinear coordinates,
\begin{subequations}
\begin{align}
&{\bf j}= (j_s, j_\nu )=\left( a \psi^*\psi, \frac{i}{2} \left\{
\psi_\eta \partial_\nu
\psi_\eta^* - \psi_\eta^* \partial_\nu \psi_\eta \right\} \right), \\
&\partial_s j_s + \partial_\nu j_\nu =0.
\end{align}
\end{subequations}

For truly stable orbits the wave functions are exponentially
localized, resulting in a minimal leakage of the probability from
the orbit boundary layer.  For unstable orbits the localization is
given by a power law, following  from the asymptotic expansion in
Eq.\ (\ref{eq:psi-as0}); the ABL wave functions then are  not
square integrable, and  the escape rate increases logarithmically
with the system dimensions. The symmetric separatrix solution
(\ref{eq:psi-unst}) combines a large maximal value in the vicinity
of the orbit with a small transverse probability current. At $\eta
<0$ the maximum of Eq.\ (\ref{eq:psi}) shifts from the orbit,
eventually moving out of the boundary layer. At $\eta>0$ the
transverse profile of the solution becomes flat and its weight
shifts from the main trajectory. [Interestingly, the separatrix
solution is also best behaved in the vicinity of focal points
(where $z\bar z=0$); for $\eta \neq 0$ the ABL solutions display a
jump by a factor $\exp ( \eta \pi)$.] The dominance of the
separatrix solution therefore has  a simple physical origin: the
$\eta$-dependent term in the quantization condition
(\ref{eq:quant-unst}) controls the transverse momentum of the
solutions, which in turn determines the leakage out of the
boundary layer.

\section{\label{sec5}Application to a quantum resonator model}

In order to assess the predictive power of the above ABL formalism
for the description of individual scarred eigenfunctions we now
turn to a specific system: a two-dimensional strip resonator
(quantum wire) in a longitudinal confinement potential and a
perpendicular magnetic field $B$.  In the $x$ direction, the
system is confined by two parallel impenetrable walls at $x=0$ and
$x=d$, while in the $y$ direction the confining potential is
parabolic, $U(y) = m \omega_0^{2}y^{2}/2$.  In the computations we
set $d = 10\, l_B$ and $\omega_0 = 2\,\omega_c$, where
$\omega_c=|e B|/m$ and $l_B=\sqrt{\hbar/m\omega_c}$ are the
cyclotron frequency and the Landau magnetic length, respectively.

\begin{figure*}[tbp]
\epsfig{file=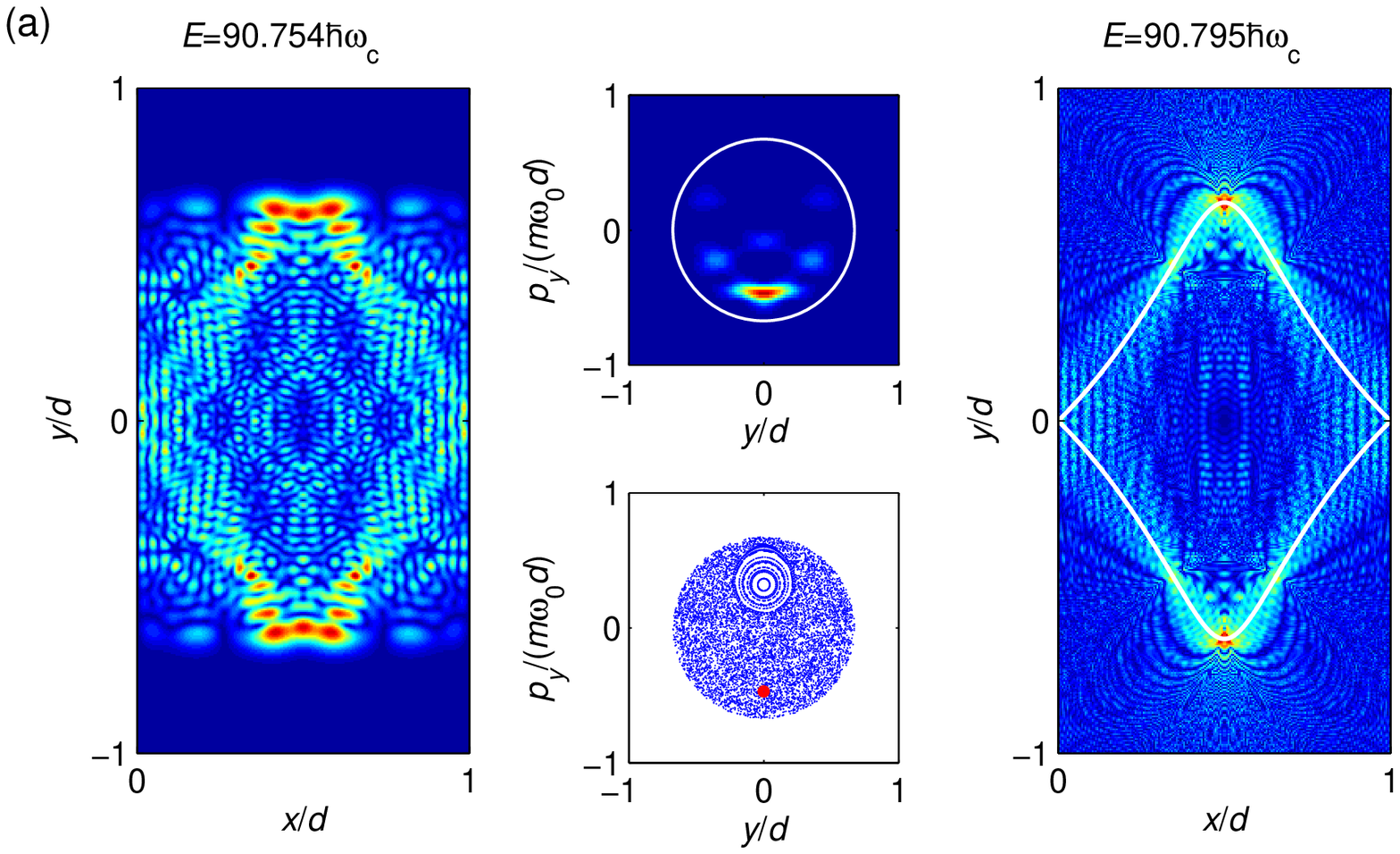,width=\columnwidth}
\hfill\epsfig{file=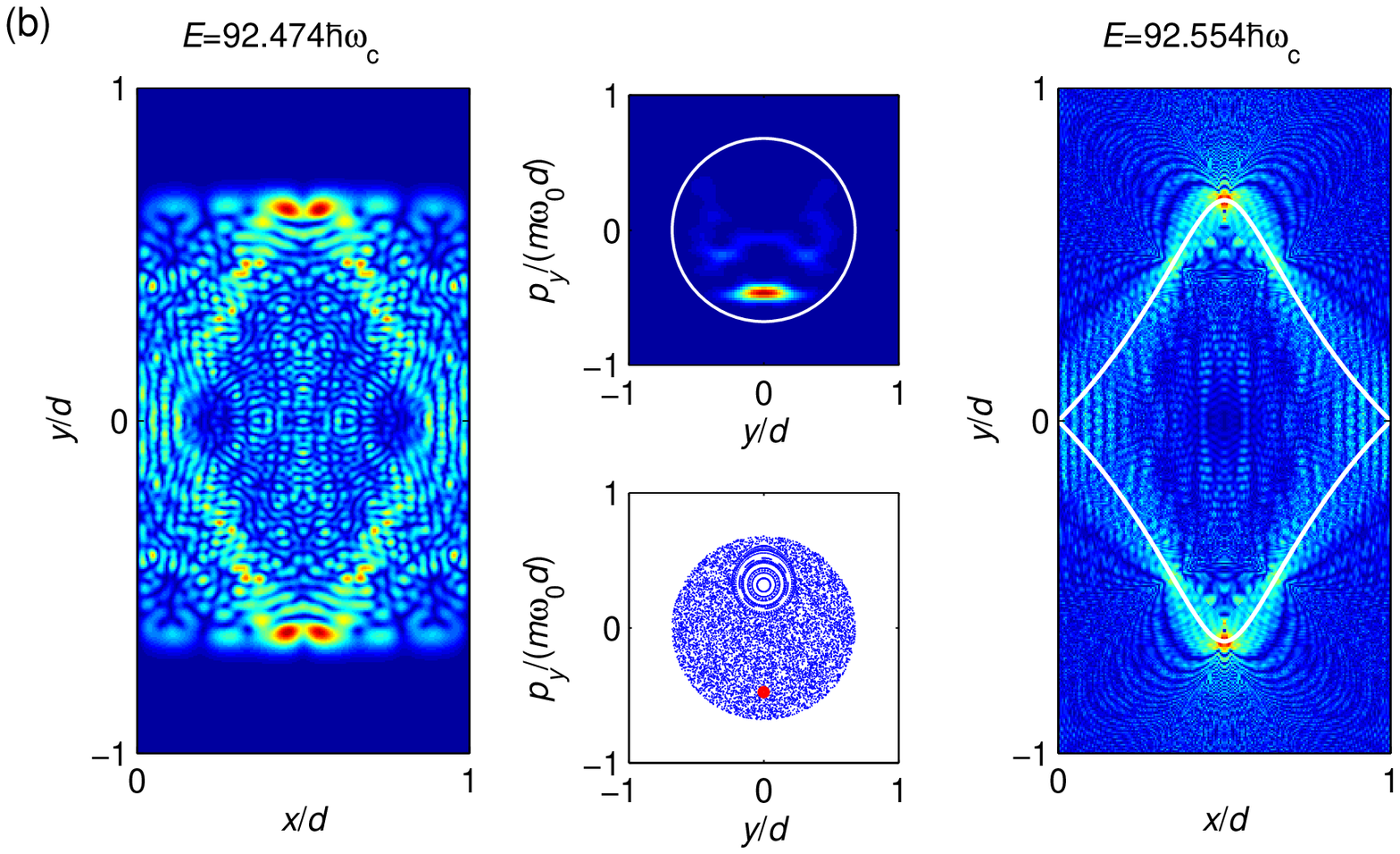,width=\columnwidth}
\epsfig{file=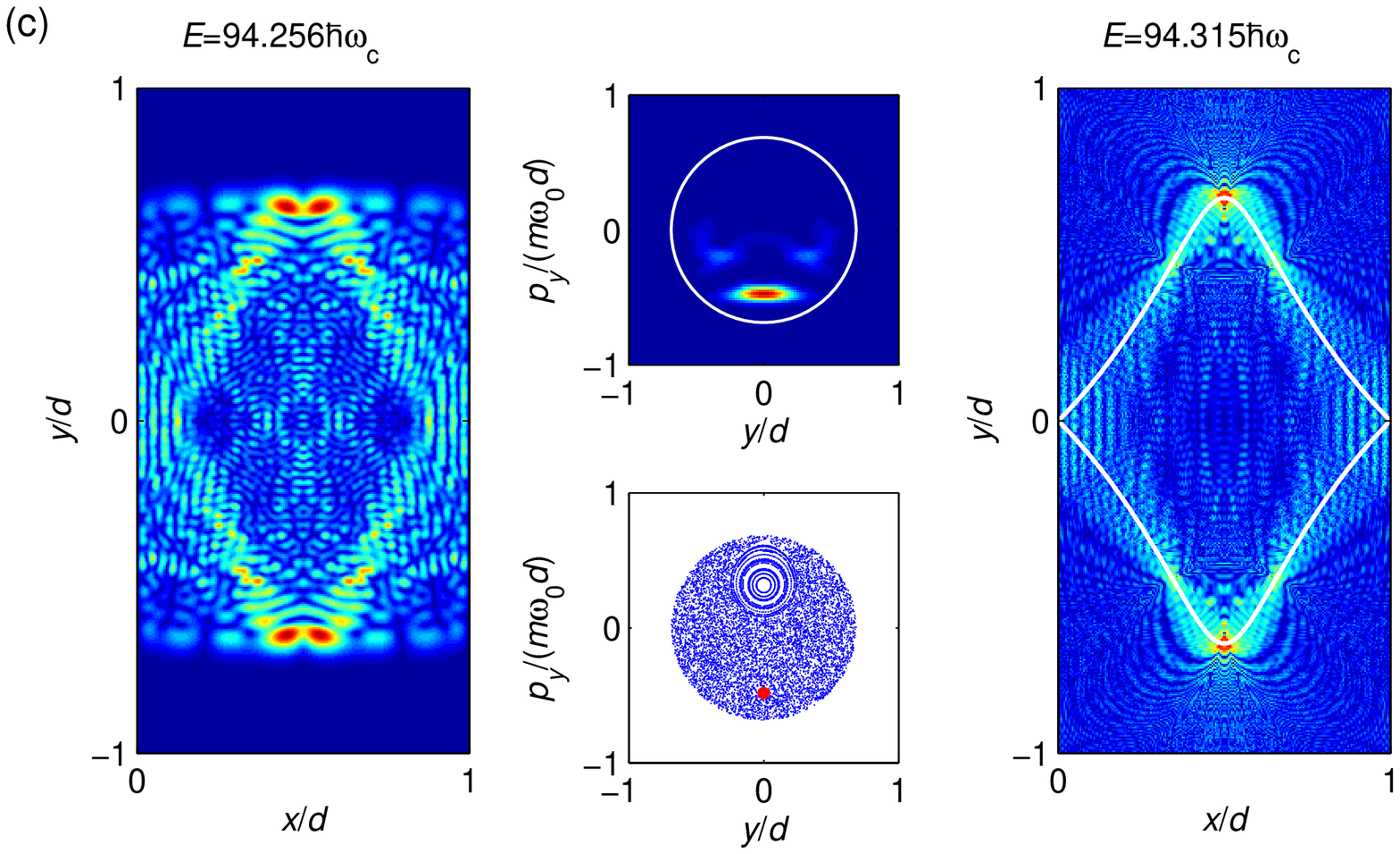,width=\columnwidth}
\hfill\epsfig{file=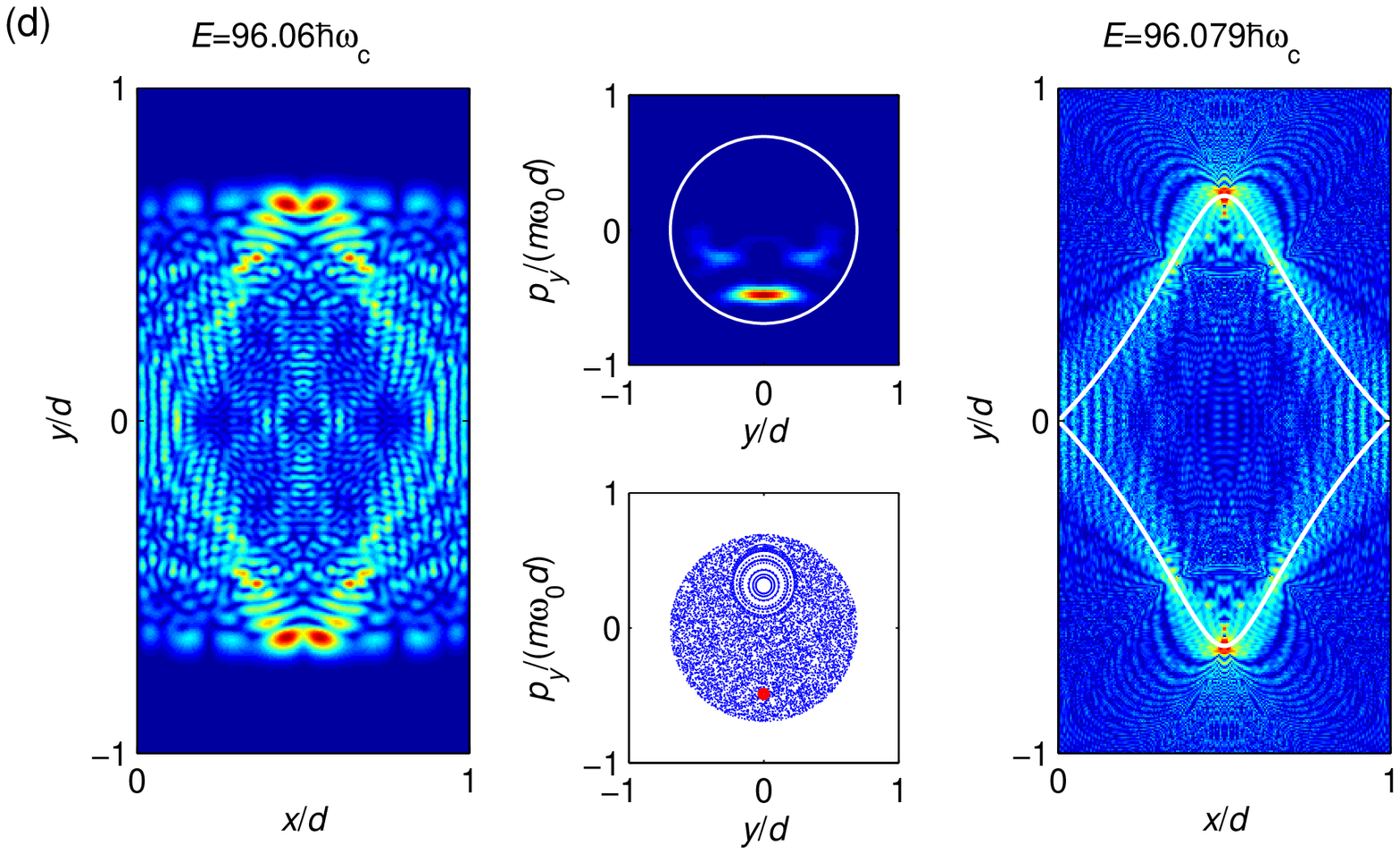,width=\columnwidth} \caption{(Color).
(a)-(d): Sequence of scarred wavefunctions localized around the
bell-shaped unstable orbit in an electronic strip resonator with a
parabolic longitudinal potential and a transverse magnetic field.
Left panels: Color-coded modulus of the exact wavefunctions
$|\Psi(x,y)|$ (blue: small amplitude, red: large amplitude). Right
panels: The corresponding semiclassical ABL wavefunctions. The
white curves depict the bell-shaped orbit, computed at the
semiclassical energy.  Upper middle panels: color-coded Husimi
representations of the exact wavefunctions.  Lower middle panels:
classical phase space portraits, computed at the energy of the
exact wavefunctions.  The red dot denotes the position of the
bell-shaped orbit.  The white circle in the Husimi representations
delimits the energetically accessible region of the classical
phase space.  } \label{fig:1}
\end{figure*}

The classical dynamics of this system is of the generic, mixed,
type, where stable orbits coexist with unstable orbits [see phase
space portraits in the middle panels of Figs.\
\ref{fig:1}(a)-(d)]. It was demonstrated previously for a similar
system \cite{Zalipaev} that the ABL method provides an accurate
description of quantum states associated with stable orbits.  Here
we analyze an energy-dependent family of unstable bell-shaped
orbits with two reflection points $(0,0)$ and $(d,0)$ at the hard
walls [see white curves in the right panels of Figs.\
\ref{fig:1}(a)-(d)].  The orbits consist of an upper and a lower
arc, which are symmetrical to each other and are indexed in the
following by $i=1,2$.

We solve the equation in variation (\ref{eq:HE}) on each arc and
then link these solutions together using a reflection matrix,
which for non-vanishing magnetic field takes the form
\begin{align}
    \label{eq:refl}
 R = \left( \begin{array}{cc} -1 & 0 \\ - 2 \omega_c \tan \Theta & -
 1 \end{array} \right),
\end{align}
where $\Theta$ is the angle of reflection.  The monodromy matrix
of the full orbit is found from the product $M = R M_2 R M_1$,
where $M_i$ is the fundamental matrix of  arc $i$. There are eight
singular points, where $z \overline z = 0$: four well isolated
focal points around $x\approx (0.5\pm 0.15)\,d$, $y\approx \pm
0.5\, d$, and two pairs of closely spaced focal points around
$x\approx(0.5\pm 0.01)\,d$, $y\approx \pm 0.65\,d$. The stretching
factor for the orbit (the largest eigenvalue of the monodromy
matrix) is $\Lambda \approx 4.6$, corresponding to $\lambda\approx
1.53$.

\begin{figure}[tbp]
\epsfig{file=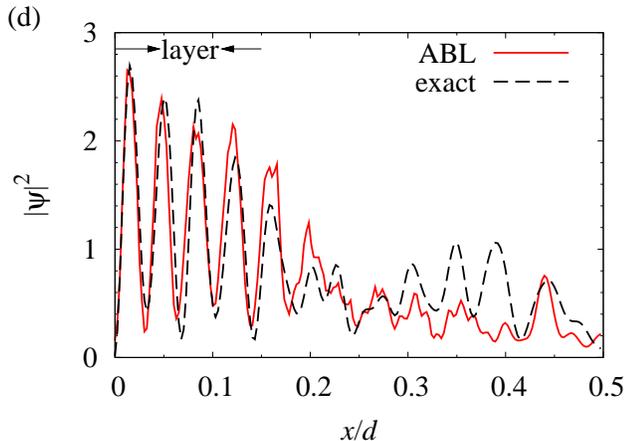,width=\columnwidth} \caption{Profile of the
exact and semiclassical wave functions calculated along the line
$y=0$.  The width of the boundary layer where the ABL approach is
valid is indicated in the upper left corner.} \label{fig:2}
\end{figure}

Semiclassically, the eigenstate is found as the sum of the two
separatrix solutions (\ref{eq:psi-unst})  along each arc,
\begin{align}
 \Psi = \psi_1(z_1,s, \nu) + r \psi_2(z_2,s,\nu).
\end{align}
Here the coefficient $r$ follows from the continuity condition at
the reflection points, and simply takes the value $r=-1$ for
unstable orbits. The condition $\Psi=0$ at the boundaries $x=0$
and $x=d$ of the resonator can be taken into account by adopting a
mirror reflection method \cite{BaKi,BaBu}. In this method, the
wave function with the required boundary condition is obtained as
the difference between the original solution and its mirror
reflection at the boundary. In principle, the presence of the
magnetic field requires to introduce an additional gauge field to
eliminate the phase jump associated to the difference
$\Gamma_{2}-\Gamma_{1} = 2 \omega_c$ entering the off-diagonal
element of the reflection matrix Eq. (\ref{eq:refl}). However,
this affects only a small neighborhood $\propto \sqrt{\hbar}$ of
the point of reflection.

In order to compare the resulting ABL solutions with  exact
quantum mechanics, we construct numerical eigenstates in an
orthogonalized basis spanned by the exact solutions of the
quasi-one dimensional system without the hard walls, which
separates in the gauge ${\bf A}=- B y {\bf e_x}$. The boundary
conditions $\Psi(x=0,d;y)=0$ are imposed via a singular value
decomposition (for details on such methods see, e.g., Ref.
\cite{numerics}).

Our numerical computations reveal that the family of bell-shaped
unstable orbits supports a long sequence of scarred wavefunctions,
which are found at almost equidistantly spaced energies.  Four
consecutive examples (with longitudinal quantum numbers
$n=66,67,68,69$) are shown in the left panels of Figs.\
\ref{fig:1}(a)-(d). The corresponding ABL wavefunctions (right
panels) accurately capture the typical spatial extent of the scar
signature, including the position of the focal points, where the
scar structure shrinks while the amplitude is significantly
enhanced (close to the almost-degenerate pairs of focal points the
wave function could be further regularized by adapting the general
techniques of uniform approximations; see, e.g., Refs.\
\cite{Berry3,bifurcations}).

The clear correspondence between the exact quantum states and the
ABL separatrix solutions  is further confirmed by the Husimi
representations $H(y,p_y)=|\langle
y,p_y|\partial_x\Psi(x=0,y)\rangle|^2$, which are obtained by
overlapping the derivative of the exact wavefunction at the left
wall with coherent states (minimal-uncertainty wave packets) that
are parameterized in Birkhoff coordinates $y,p_y$. The Husimi
representation, shown in the upper middle panels of Figs.\
\ref{fig:1}(a)-(d), in all cases displays a clear maximum at the
position of the bell-shaped orbit (red dot in the chaotic part of
the phase-space portraits), which lies in the chaotic part of
phase space.

Besides the visual agreement, the separatrix solutions also
recover the following essential characteristics of the numerical
solutions: (i) The wavefunctions are of the right symmetry with
respect to inversion around $x=0.5\,d$, $y=0$ (as there are eight
focal points, wavefunctions are symmetric under this operation
when $n$ is odd, while they are antisymmetric when $n$ is even).
(ii) The profile of the wave function along the line $y=0$
extracted from the ABL and the numerical solution (after spatial
averaging over a range of $0.05\,d$ to eliminate the speckle
fluctuations due to the random background  in the numerical
solution) shows a very good agreement within validity region of
the ABL solution, $x/d \lesssim 0.15$. The ABL captures well both
the period and the relative strength of the oscillations in the
numerical solution. (iii) The exact energies are in excellent
agreement with $E_{n_0}$, i.e., the semiclassical prediction from
Eq.\ (\ref{eq:quant-unst}) with $\eta=0$. These energies are
almost equidistantly spaced, and the semiclassical error is less
than $5\%$ of this spacing (the error is also small compared to
the mean level spacing of all states). (iv) For each longitudinal
wavenumber $n$ the numerical computations only deliver a single
scarred wavefunction localized around the unstable trajectory (in
contrast, many transversely excited wavefunctions are supported by
the stable trajectories in the island around $y\approx 0$,
$p_y\approx 0.3\,m\omega_0 d$). In other words, we do not find any
other scarred states on this orbit. Therefore, the actual scar
quantization window is reduced beyond the constraint obtained by
the propagation of time dependent Gaussian states \cite{Heller3},
which predicts scarring in a larger window $\Delta E \approx \hbar
\lambda_T$. However, one cannot exclude that signatures of
off-resonant scarring show up in the wavefunction statistics of
the states in this larger window, which is a question beyond our
focus on individual wave functions.

\section{\label{sec6}Conclusions}

In this work we extended the asymptotic boundary layer (ABL)
method for semiclassical wave functions, originally developed for
the description of Gaussian beams guided by stable trajectories,
to the case of scar-like states localized in the vicinity of
unstable periodic orbits. We focus on wavefunctions scarred by a
single short trajectory and derive expressions of a universal
form, valid up to order $\hbar^{3/2}$, which apply to general
systems which may combine hard walls, external potentials and
magnetic fields. The ABL equations are formulated in a curvilinear
coordinate system associated with the classical trajectory. The
system- and orbit-specific information enters via the solution of
the classical equations in variations, which describe the
stability of the trajectory. At fixed energy, the profile of the
wavefunction is determined by a periodicity condition, which
results in a Bohr-Sommerfeld quantization formula.

Far away from the guiding trajectory, the ABL wave function decays
as the inverse square-root of the distance, and therefore is not
normalizable, but couples to the quasi-random background of other
modes in the system.  This coupling is the weakest for the
symmetric separatrix state, which is associated to a particulary
simple quantization condition. Since the separatrix state is also
the most visibly localized ABL wavefunction, one can expect that
it typically provides the dominant  contribution to scars guided
by a single short trajectory. In general, we estimate that the
number of wave functions of different profile participating in the
scar formation increases logarithmically with the system
dimensions, and therefore depends on the relation  of the
Ehrenfest time and the period of the trajectory.

We verified our conclusions by  comparison with  exact numerical
solution for a quantum strip resonator with the quadratic
confining potential and a perpendicular magnetic field. In this
system, a family of unstable periodic orbits supports a long
sequence of scarred states, with energies in excellent agreement
with the simple quantization condition for the separatrix
solution. We find that this solution captures all the essential
characteristics of the exact scarred states. The studied system is
of the generic type, with a mixed phase space, which raises the
expectation that the ABL formalism is applicable to a large class
of quantum-dynamical systems.

An open question concerns the generalization of the formalism to
scars supported by many trajectories, as well as the contribution
of ABL solutions to off-resonant scars (typically revealed in
wavefunction statistics).

This work was supported by the European Commission via Marie-Curie
excellence grant No. MEXT-CT-2005-023778.

\begin{appendix}
\section{ABL expansion for Schr\"odinger equation}

In this appendix we outline the main steps in the derivation of
the ABL formalism. Further details on the formalism can be found
in Refs.\ \cite{BaKi,BaBu}. Here, we present a straightforward
derivation based on the direct expansion of the Schr\"odinger
equation.

The main trajectory ${\bf r}_{0}(s)$, which is parameterized by
its arclength $s$, generates an orthogonal coordinate system
spanned by the normal unit vector ${\bf e}_n(s)$ and the
longitudinal unit vector $ {\bf e}_t(s)$, which we express in the
component form
\begin{align}
 {\bf e}_n(s) \equiv \left( \begin{array} {c} \gamma_x \\ \gamma_y
\end{array}
\right), ~ {\bf e}_t(s) \equiv \left( \begin{array} {c} \gamma_y \\
-\gamma_x \end{array} \right).
\end{align}
Recalling the conversion  formulas for the partial derivatives in
curvilinear coordinates,
\begin{align}
&\partial_x = \frac{\gamma_y }{\xi} \partial_s + \gamma_x \partial_n,
~ \partial_y = - \frac{\gamma_x }{\xi} \partial_s + \gamma_y
\partial_n, \\
&\xi = 1-\frac{n}{\rho(s)}, \notag
\end{align}
where $\rho(s)$ is the curvature radius of the trajectory, we
write the Schr{\"o}dinger equation as
\begin{align}
&\frac{\Psi_{ss}}{\xi^2} - \frac{n \dot{\rho }}{\xi^3 \rho ^2} \Psi_s
-\frac{\Psi_n }{\xi \rho } + \Psi_{nn} + \frac{i {e B}}{\hbar \xi}
\left[ ({\bf r}_0 {\bf e}_n) + n \right] \Psi_s \notag \\
&+ \frac{i {e B}}{\hbar } ({\bf r}_0 {\bf e}_l ) \Psi_n -
\frac{{e^2 B}^2} {4 \hbar^2 \xi^3}( {\bf r}_0 + {\bf e}_n n)^2 \Psi =
\frac{2m}{\hbar^2 \xi^3} (U - E) \Psi,
\label{eq:SCHROD}
\end{align}
where $U(s,n)$ is the potential
inside the resonator.

The central idea of the ABL approach is to search the solution to
Eq.\  (\ref{eq:SCHROD}) in an asymptotically small boundary layer
of the width $\propto \sqrt{\hbar}$ in the vicinity of the main
trajectory. A straightforward way to derive the corresponding
semiclassical expansion for the wave function follows two steps:
(i) scaling the normal variable as $n = \sqrt{\hbar} \nu$ and (ii)
expanding the resulting equation in powers of $ \sqrt{\hbar}$, as
written in Eq. (\ref{eq:WKB}), where each term $\psi^{(j)}$ is
uniquely defined.

We substitute Eq.\ (\ref{eq:WKB}) into Eq.\  (\ref{eq:SCHROD}),
expand the resulting equation, and match the coefficients of the
resulting series, which yields a hierarchical set of equations for
$S_{0}(s)$, $S_{1}(s)$, and $\psi^{(j)}(s,\nu)$.  If the expansion
of the effective potential can limited to second order (a standard
assumption in  linear scar theories \cite{Kaplan,Dambrowsky}), one
can terminate the series in Eq.\ (\ref{eq:WKB}) at the leading
term $\psi^{(0)}(s,\nu)$. This is equivalent to obtaining a
semiclassical wave function to the order of $\hbar^{3/2}$
\cite{BaKi}. The first three terms of the series expansion of Eq.\
(\ref{eq:SCHROD}) read
\begin{subequations}
\begin{align}
\label{eq:1}
& \hbar^{-2}: & {\cal A} \ \psi^{(0)} = 0, \\
\label{eq:2} & \hbar^{-3/2}: & {\cal A} \ \psi^{(1)} + 2 i {\cal
B}\ \psi^{(0)}_\nu +
{\cal C} \nu \ \psi^{(0)} = 0, \\
\label{eq:3} &\hbar^{-1}: & {\cal C} \nu \ \psi^{(1)} + 2i {\cal
B}\ \psi^{(1)}_\nu + {\cal D} =0,
\end{align}
\end{subequations}
with coefficients
\begin{subequations}
\begin{align}
{\cal A} &= a^2 - \frac{r_0^2 {e^2 B}^2}{4} - {e B} ({\bf r}_0
{\bf
 e}_l) S_1 - S_1^2 + {e B} ({\bf r}_0 {\bf e}_n) {\dot S}_{0}
 - {\dot S}_{0}^2,  \\
{\cal
B} &= \frac{e B}{2} ({\bf r}_0 {\bf e}_l)+S_1,  \\
\label{eq:coeff-C}
{\cal C} &= \frac{S_1^2 - {\dot S}_0^2}{\rho} - 2 {\dot S}_0{\dot
S}_1 - e B {\dot S}_0^2 -
e B {\dot S}_1 ({\bf r}_0 {\bf e}_n)  \notag \\
&- \frac{eB}{\rho} ({\bf r}_0 {\bf e}_l) S_1 - 2m u_1 - \frac{a^2}{\rho} + \frac{e^2 B^2}{4\rho} r_0^2 - \frac{e^2 B^2}{2} ({\bf r}_0 {\bf e}_n), \\
{\cal D}&= 2 i a\ {\dot \psi}^{(0)} + i {\dot a}\ \psi^{(0)} +
\psi^{(0)}_{\nu \nu} - d a^2 \nu^2 \ \psi^{(0)} .
\end{align}
\end{subequations}

Equation  (\ref{eq:1}) yields ${\cal A} =0$. Equation (\ref{eq:2})
is solved by setting ${\cal B}=0$, since ${\cal C}=0$ is satisfied
identically if ${\cal A} ={\cal B} =0$, which yields the classical
action solution (\ref{eq:S}). Finally Eq. (\ref{eq:3}) is
equivalent to ${\cal D}=0$, which yields Eq. (\ref{eq:SL}). After
the substitution $\phi = a^{-1/2} \psi^{(0)}$ this equation take
the form of a nonstationary Schr\"odinger equation, which we write
as
\begin{align}
\label{eq:SL1}
    \hat L \phi \equiv \left\{ i \partial_s + \frac{\partial_\nu^2 }{2
    a}- \frac{a d}{2} \nu^{2} \right\}\phi = 0.
\end{align}

The fact that Eq.\ (\ref{eq:psi}) is a solution to Eq.\
(\ref{eq:SL}) can be checked by a direct substitution. Here we
present a direct derivation, which  provides deeper insight into
the nature of the ABL solution. First, we obtain a ``ground
state'' solution to Eq. (\ref{eq:SL1}) by using the ansatz
\begin{align}
\label{eq:GR}
 \phi_0 = \mu(s) \exp \left\{ i \ \frac{ \Gamma(s)}{2}\ \nu^2
 \right\},
\end{align}
which is similar to the thawed Gaussians introduced by Heller
\cite{Heller_g} to describe the time evolution of localized
quantum wave packets.  Substituting Eq.\ (\ref{eq:GR}) into Eq.
(\ref{eq:SL1}) yields equations for the unknown quantities
$\Gamma$ and $\mu$,
\begin{subequations}
\begin{align}
\label{eq:RIC}
&\dot \Gamma + \frac{\Gamma^2}{a} + a d =0, \\
\label{eq:mu}
&\mu = \exp \left\{ - \frac{1}{2} \int_0^s
\frac{\Gamma}{a} ds \right\}.
\end{align}
\end{subequations}
The Ricatti equation (\ref{eq:RIC}) is solved by a standard
substitution $\Gamma = p/z$, after which one finds equations for
$p$ and $z$ as given by Eq.\ (\ref{eq:HE}). As we mentioned above,
this Hamiltonian {\em equations in variations} define relative
coordinates $z$  and momentum $p$ for  classical trajectories in
the vicinity of the guiding trajectory. Once Eq.\ (\ref{eq:HE}) is
solved,  Eq.\ (\ref{eq:mu}) delivers $\mu = z^{-1/2}$. Equation
(\ref{eq:HE}) has two linearly independent solutions, $(z,p)$ and
$({\overline z},{\overline p})$, which define the constant
Wronskian $p {\overline z} - {\overline p} z = w$.

The classical origin of Eq.\ (\ref{eq:HE}) can be established from
the reduced classical action constructed for the trajectory $n(s)$
in the vicinity of the guiding trajectory. In curvilinear
coordinates this action reads
\begin{align}
&S = \int_0^s \sqrt{2m (E- U(r))} \left\{ \left(1- \frac{n}{\rho} \right)^2
+ {\dot n}^2 \right\}^{1/2} ds \notag \\
&+ \frac{e B}{2} \int_0^s \left( x \dot y - y \dot x \right) ds,
\label{eq:S_short}
\end{align}
where $x = x_0(s) + \gamma_x(s) n(s)$ and $y = y_0(s) +
\gamma_y(s) n(s)$.  Expanding Eq.\  (\ref{eq:S_short})  to second
order in the small quantity $n$, one obtains
\begin{align}
\label{eq:S_exp}
&S = S_0 - \int_0^s  \varepsilon \ n \ ds + \frac{1}{2} \int_0^s \left( a \dot
n^2 - a d n^2 \right) ds,
\end{align}
where
\begin{align}
 \varepsilon(s) = {e B} + \frac{a} {\rho} + \frac{a u_1} {2(E -
 u_0)}.
\end{align}
The linear term in this expansion vanishes identically,
$\varepsilon=0$, because $n=0$ is also a trajectory (the main
trajectory). The linear coefficient in Eq. (\ref{eq:S_exp}) is related to Eq. (\ref{eq:coeff-C}) as ${\cal C} = 2 a \varepsilon$, which explains why ${\cal C}$ vanishes.  The Euler-Lagrange equation for the action
(\ref{eq:S_exp}) is equivalent to the Hamiltonian system
(\ref{eq:HE}).

A sequence of partial solutions to Eq.\  (\ref{eq:SL}) can be
constructed with the help of creation and annihilation operators
\cite{Mas1,BaKi}
\begin{subequations}
\begin{align}
&\hat \Lambda =  \sqrt{\frac{i}{w}} \left( -i \ z\  \partial_\nu - p\ \nu
\right), \\
&\hat \Lambda^\dagger =  \sqrt{\frac{i}{w}} \left( - i \ {\overline z}\
\partial_\nu - {\overline p}\ \nu \right),
\end{align}
\end{subequations}
which satisfy the following algebra
\begin{align}
[\hat \Lambda,\hat L] = [\hat \Lambda^\dagger,\hat L] = 0, \ [\hat
\Lambda,\hat \Lambda^\dagger] = 1.
\end{align}
General solutions are then obtained from $\psi_m =
\Lambda^{\dagger m} \psi_0$. This procedure leads to a recurrence
relation, which is solved by  wave functions of the explicit form
\begin{equation}
    \label{eq:psi1}
    \phi_{m}(s,\nu) \propto \frac{ {\overline z}^m (\Gamma-{\overline
    \Gamma})^{m/2}} {\sqrt{z}} H_{m} \left( \sqrt{\frac{\Gamma -
    {\overline \Gamma}}{2i}} ~\nu \right) e^{\frac{i}{2} \Gamma \nu
    ^{2}},
\end{equation}
where $H_{m}$ are Hermite polynomials.

The presented derivation allows for an unambiguous analytical
continuation of the recurrence relation and its solutions to
arbitrary indices $m \rightarrow \xi$. This yields Eq.\
(\ref{eq:psi}), where the parabolic cylinder functions $D_\xi$
take the place of the Hermite polynomials $H_m$, and the
coefficient $a^{-1/2}$ is accounted for. Finally, we note that
recalling the definition of $\Gamma$ and $\overline \Gamma$ as
well as that for the Wronskian, the solution can be compactly
written as
\begin{equation}
    \label{eq:psi2}
    \phi_{\xi}(s,\nu) \propto \frac{e^{\frac{i}{4} (\Gamma+\overline \Gamma ) \nu ^{2}} }{\sqrt{z}}
    \left(\frac{\overline z}{z} \right)^{\xi/2} D_{\xi} \left(
    \sqrt{\frac{w}{i z\overline z }} ~\nu \right).
\end{equation}

\end{appendix}

\end{document}